\documentclass{arxivversion}
\usepackage{cite}
\usepackage{amsmath,amssymb,amsfonts}
\usepackage{algorithmic}
\usepackage{graphicx}
\usepackage{textcomp}
\usepackage{comment}

\usepackage{amsmath}
\usepackage{nccmath}
\usepackage{multicol}
\usepackage{multirow}
\usepackage{makecell}
\usepackage{array}
\usepackage{url}

\def\BibTeX{{\rm B\kern-.05em{\sc i\kern-.025em b}\kern-.08em
    T\kern-.1667em\lower.7ex\hbox{E}\kern-.125emX}}
\begin{document}
\doi{}

\title{Accurate smartphone camera simulation using 3D scenes}
\author{
\uppercase{Zheng Lyu, Thomas Goossens, Brian Wandell and Joyce Farrell}}
\address{Stanford Center for Image Systems Engineering (SCIEN), Stanford, CA 94305}

\tfootnote{This work was supported in part by funds donated to the Stanford Center for Image Systems Engineering by Google}

\markboth
{\headeretal: Smartphone camera simulation}
{\headeretal: Smartphone camera simulation using 3D scenes}

\corresp{Corresponding author: Zheng Lyu (e-mail: zhenglv.felix@gmail.com).}

\begin{abstract}
We assess the accuracy of a smartphone camera simulation. The simulation is an end-to-end analysis that begins with a physical description of a high dynamic range 3D scene and includes a specification of the optics and the image sensor. The simulation is compared to measurements of a physical version of the scene. The image system simulation accurately matched measurements of optical blur, depth of field, spectral quantum efficiency, scene inter-reflections, and sensor noise. The results support the use of image systems simulation methods for soft prototyping cameras and for producing synthetic data in machine learning applications. 
\end{abstract}


\titlepgskip=-15pt

\maketitle
\section{Introduction} \label{Intro}
Computer simulation is a powerful tool for modeling and evaluating scientific ideas and engineering solutions in many fields. Simulations are often used by engineers to design and test individual components of an imaging system, including lenses and sensors. Image system simulations model how properties of the scene, optical components, and sensor interact to produce the final sensor image. These end-to-end simulations can be used to design rendering methods for consumer photography or to develop machine learning algorithms \cite{jiang2017learning, liu2021isetauto}. 

Together with our colleagues, we implemented image systems simulation software to model the complete image processing pipeline of a digital camera, including scene radiance, image formation by the optics, sensor capture, image processing and display rendering.  This work, initiated in 2003, was developed to support the design of digital imaging sensors for consumer photography and mobile imaging \cite{Farrell2012-ma, Farrell2003-rb}. The original simulations were validated by comparing simulated and real digital camera images using calibrated 2D test charts \cite{Farrell2008-sc, Chen2009-bj}. 

In recent years, we added the ability to model  three-dimensional scenes into the image systems simulation \cite{Liu2019-eo}. Using physically based ray tracing \cite{Pharr2016-yb}, we calculate the spectral irradiance image of complex, high-dynamic range (HDR) 3D scenes. This extension opens the opportunity for accurately simulating natural scenes with complex lighting, high dynamic range, occluding objects, and surface inter-reflections. 

We have used three-dimensional scenes to design and evaluate imaging systems for a range of applications, including AR/VR \cite{Lian2018-kc}, underwater imaging \cite{Blasinski2017-lr}, autonomous driving \cite{Blasinski2018-em, Liu2019-ds}, fluorescence imaging \cite{Farrell2020-tb, Lyu2021-ae} and models of the human image formation \cite{Lian2019-xr}. In each case, we carried out certain system validations - typically by comparing some aspect of the simulation with real data.   In this work, we validate the software by comparing camera images measured using a constructed Cornell Box \cite{Goral1984-wd, Meyer1986-pb} with simulations of the scene and camera (Figure \ref{Fig-scenepreview}).  The box contains one small and one large rectangular object placed near the side walls of the box. The box is illuminated through a hole at the top, so that some interior positions are illuminated directly and others are in shadow. The walls of the box are covered with red and green paper, providing colorful indirect illumination of the rectangular objects. 

The Cornell Box was designed to include important features found in complex natural scenes, such as surface-to-surface inter-reflections and shadows. In its original use, the Cornell Box was used to judge the visual similarity between a real constructed box and computer graphics renderings of the box displayed on a CRT. Ray tracing algorithms capture the main effects of the lighting and inter-reflections \cite{Goral1984-wd, Cohen1986-ld, Meyer1986-pb,Sumanta-qr}.  
Our study is designed to assess the accuracy of end-to-end simulations that include a physical description of a 3D spectral radiance, realistic optics, and an electronic image sensor. We validate the end-to-end simulations -- from the Cornell box to the sensor digital values -- by comparing simulated and real camera images. 

\section{Related work} \label{relwork}

A number of groups have recognized the importance of image systems simulation using the principles of optics and sensor simulations \cite{Farrell2003-rb, Costantini2004-mn,Konnik2014-mu,Toadere2013-au, florin2009simulation, fiete2014modeling}. There are many commercial offerings that provide engineers with simulation tools for a range of applications (e.g., dSpace, ANSYS, Anyverse, Omniverse-Nvidia, Zemax, CodeV).

Among the first systems were those developed for remote sensing.  Schott et al. \cite{schott1999advanced} developed image systems software (DIRSIG) that includes radiometric descriptions of scenes, ray tracing through atmospheric media, and sensor capture. The goal of the software is to create synthetic broad-band, multi-spectral and hyper-spectral imagery. Some aspects of the simulations are validated \cite{White1996-lv, Peterson2004-la, dirsig-validation-se}.  The code is distributed in binary form to those who pay to attend a course and who hail from a specific set of countries. A large number of related papers describing in-house simulations have also been reported (e.g., \cite{GISGeography2014-ez,Borner2001-cn}).

Garnier et al \cite{garnier1998general, garnier1999infrared} describe an end-to-end image systems simulation of an infrared (IR) sensor model.  Their simulation traces rays from a scene, through an optics model, and to a sensor array. The optics model combines both thin-lens equations and blurring kernels applied to the final image. The computational concepts are very similar to the methods we use, and their paper is a good overview of the physics of image formation and sensor capture. No experimental validation of this work is published and we were unable to locate the software.

Image systems simulations have been a powerful tool in our research and have enabled us to build soft-prototypes of imaging sensors before they were built. For example, we used simulations to develop the L\textsuperscript{3} (local, linear and learned) method for processing RGBW sensor data \cite{Lansel:2011, Tian:2014, Tian:2015, Jiang:2017}.  We also used simulations to design a spectrally-adaptive imaging system for underwater imaging \cite{blasinski2017computational} and a camera that is capable of measuring the fluorescence of oral bacteria \cite{Farrell2020-tb, Lyu2021-ae}.  

More recently, we used simulations to compare the performance of neural networks for detecting vehicles in complex daytime scenes. In one study, we explored the effect that sensor pixel size has on performance \cite{liu2020neural}. And, in another study, we compared the performance of imaging sensors that are capable of collecting radiance or depth information, and a hybrid imaging sensor that can capture both radiance and depth simultaneously on the same sensor array \cite{liu2021isetauto}.  

The ISETCam and ISET3d software we describe relies on physically based ray tracing that is capable creating quite complex natural scenes \cite{Pharr2016-yb}. The software can describe radiance in any spectral band, and it can specify a wide variety material properties including fluorescence \cite{Lyu2021-ae}. We added the capability of modeling arbitrary imaging optics specified by a ray transfer function \cite{Goossens-RTF} and thus compute the sensor irradiance.  The ISETCam software models many properties and types of imaging sensors, including sensors with arbitrary color filter arrays and sensors with microlens arrays to capture the light field.   

As far as we are aware, at this time ISETCam and ISET3d are the only open-source, freely available image systems simulation software. Using ISETCam and ISET3d, scientists and engineers can check each others' work and extend the system for their specific applications. 

\section{Contributions}
In this paper, we evaluate the accuracy of the simulations by comparing the pixel values in real camera images of a real Cornell Box with the pixel values in simulated camera images of a simulated Cornell Box. Our specific contributions include:
\begin{enumerate}
         \item Quantifying the accuracy of end-to-end image system simulations using a real 3D scene (Cornell Box)  and a real camera (Google Pixel 4a), including
         \begin{enumerate}
             \item Scene with shadows, lighting and surface inter-reflections
             \item Optical blur, relative illumination, and depth of field
             \item Sensor noise, channel cross-talk, and chromatic responses
         \end{enumerate}
\item Providing the data as well as the open-source and free software to support transparency and reproducibility
        \begin{enumerate}
            \item {https://github.com/ISET/isetcam} 
            \item {https://github.com/ISET/iset3d}
            \item {https://github.com/ISET/isetcornellbox}
        \end{enumerate}
\end{enumerate}

\section{Simulation pipeline modeling} \label{parameters}
This Section reviews the image systems simulation pipeline modeling. Section \ref{val} describes the validation measurements. Section \ref{Discussion} reviews the results. Section and \ref{connfuture} describes ongoing and future work. 

The image systems simulation validation includes three main parts: (1) a 3D scene radiance description that models light sources, asset geometry, and material properties, (2) an optical model that maps rays from the scene onto the sensor, and (3) a model of how the irradiance at the sensor is converted into unprocessed digital values. 

The validation measures how closely the simulation predicts the minimally processed sensor data obtained from the Google Pixel 4a. We focus on matching the unprocessed sensor data because remaining components of the image processing are implemented in proprietary software. 

\subsection{Scene construction}
\begin{figure}[!tb]
  \includegraphics[width=1\columnwidth]{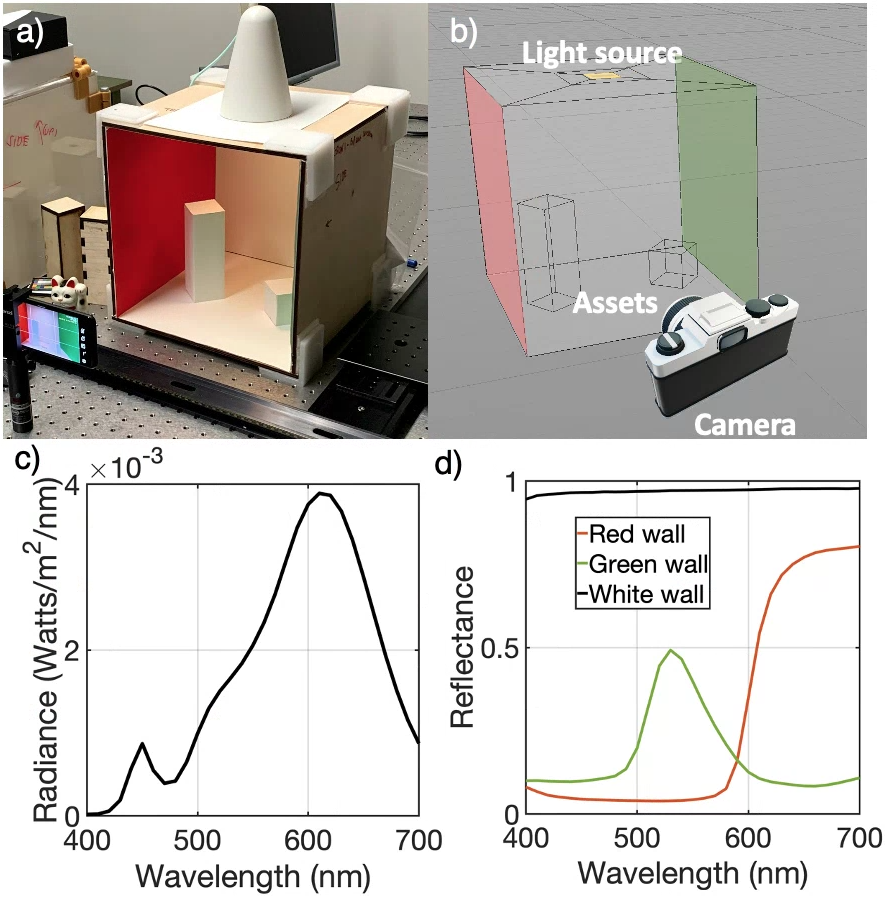}
  \caption{Real and simulated Cornell box. (a) An image of the Cornell box and its light source in the lab. (b) Geometric representation of the Cornell box model and camera position in Cinema 4D. (c) Measured spectral power distribution of light source. (d) Measured spectral reflectance of white, red and green surfaces.}
  \label{Fig-scenepreview}
\end{figure}

We constructed a Cornell Box and created a computer graphics model of the same box in our simulation environment. The Cornell Box was constructed using wood and covered with white matte paper. We added red and green matte paper to the left and right walls, respectively. We placed a light behind a diffusing filter over an opening at the top of the box (Figure \ref{Fig-scenepreview} (a)). The spectral power distribution of the light source and the spectral reflectances of the surfaces in the Cornell Box were measured using a PR670 spectroradiometer \ref{Fig-scenepreview} (b-c). In different experiments we added a miniature Gretag color chart, a slanted bar target, and a 3D printed Stanford Bunny \cite{stanford-bunny}.

The Cornell Box is a good compromise between simplicity and complexity:  it includes a light source and a set of objects within a closed environment that includes many inter-reflections and shadows. There are only a few materials, making it possible to accurately model the scene. The light and objects within the box create significant shadows and a relatively high dynamic range image. Thus, the scene is useful for testing  complex rendering, dynamic range, and  sensor noise. The surfaces with different reflectances can be used to evaluate sensor chromatic responses and surface inter-reflections. For example, we were able to assess the estimated spectral quantum efficiency of the sensor by adding the Gretag color chart to the scene. Similarly, we assessed the optics model by varying the position of the slanted bar target added to the scene. 

A Cornell Box was constructed using geometric specifications of a model of the Cornell Box represented in Cinema 4D (Figure \ref{Fig-scenepreview} (b)), a commercial graphics modelling software. The meshes and their positions are exported as a set of text files that can be read by PBRT. We  model the light source as an area light, and we model the surface materials as mainly Lambertian, but with a small specular term.  The reflectance distribution function for each surface is documented in the software we share.  The measured spectral data of the light source (Figure \ref{Fig-scenepreview} (c)) and surface reflectances (Figure~\ref{Fig-scenepreview} (d)) are stored in data files that are read into PBRT.

\subsection{Optics modelling}

\begin{figure}[!hb]
  \includegraphics[width=1\columnwidth]{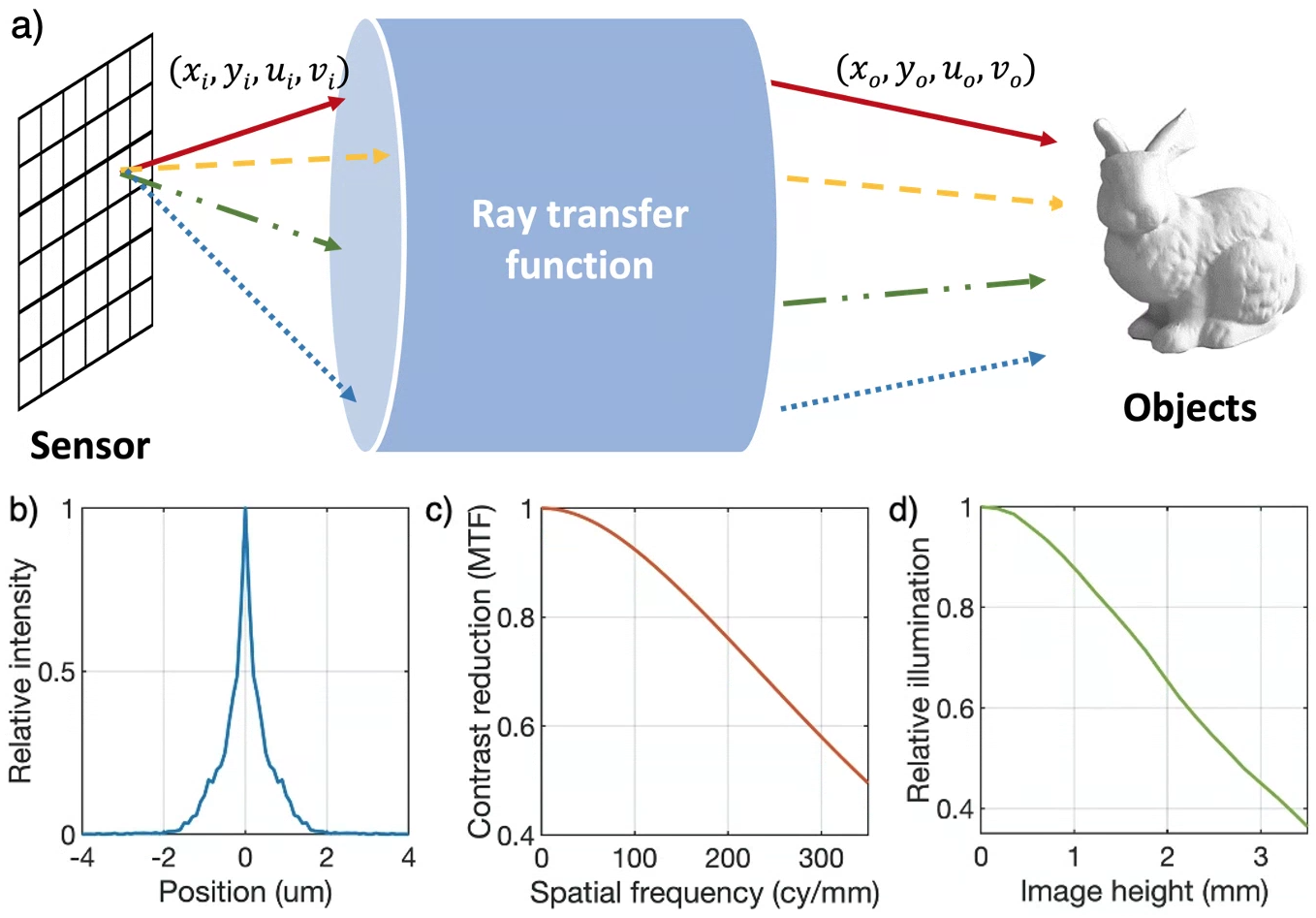}
  \caption{The ray transfer method of lens modeling. (a) We use a Zemax 'black-box model' to estimate the ray-transfer function.  This function describes the mapping between rays from the sensor plane that are incident at the entrance pupil, through the unknown optics, to rays in a plane after exiting from the optics. The position of the incident and output rays are described by the four-dimensional vectors, $(x_i,y_i,u_i,v_i)$ and $(x_o,y_o,u_o,v_o)$. The first two entries define the position of the ray in the aperture, and the second two entries define the ray direction. We use Zemax to calculate 128,561 input-output ray pairs. We then fit a set of four polynomials relating the four input ray parameters to each of the four output ray parameters (Equation \ref{eq:polynomial}). (b-d) The Zemax model generates a line spread function (LSF) , modulation transfer function (MTF) and relative illumination function, all at 550 nm.  The ray-transfer function implementation in PBRT matches the Zemax calculations on these functions \cite{Goossens-RTF}.  We evaluate the model accuracy by comparing measurements with these simulations.}
  \label{Fig-rtfzemax}
\end{figure}

Optics are a critical component of image systems and determine important properties, such as lens shading (vignetting), spatial resolution, and depth of field. It is possible to use standard ray tracing to quantify these properties when the shapes, positions, and indices of refraction of the optical components are known. In some cases, including the project analyzed in this paper, the lens design is proprietary. 

To solve this problem, we used the approach described in \cite{Goossens-RTF}, who point out that for ray-tracing purposes we only need to know the "ray-transfer function" of the optics. This function maps rays entering the optics into outgoing rays. We can simulate a proprietary lens by finding an equivalent ray-transfer function using a "black box" model provided by the vendor.  This model enables a customer to calculate how a ray in the incident light field (at a position and angle in the aperture) is mapped to a position and angle of a ray in the exit pupil, without revealing the lens design (Figure \ref{Fig-rtfzemax}).  

We use the methods in \cite{Goossens-RTF} to calculate the polynomials that describe the Google Pixel 4a ray-transfer function. Conceptually, four input ray parameters are related to four output ray parameters by four polynomials. These polynomials can be calculated for each sampled wavelength:
\begin{equation}
\label{eq:polynomial}
    \begin{cases}
        x_o =& \text{poly}_1(x_i, y_i, u_i, v_i)\\
        y_o =& \text{poly}_2(x_i, y_i, u_i, v_i)\\
        u_o =& \text{poly}_3(x_i, y_i, u_i, v_i)\\
        v_o =& \text{poly}_4(x_i, y_i, u_i, v_i)\\
    \end{cases}.       
\end{equation}
The $x,y$ values describe the position of the rays within the input or output aperture of the optics.  The $u,v$ values are the first two elements of the unit vector that describes the direction of the ray with respect to the input or output aperture.  Only two components are needed because the direction is a unit vector. As described in \cite{Goossens-RTF}, this formulation can be simplified for the common case when the optics are rotationally symmetric, and we assumed rotational symmetry here.


The ray-transfer function can be used to calculate many summary measures of the optics, including the linespread function, modulation transfer function, and the relative intensity map (Figure~\ref{Fig-rtfzemax}b-d). It is convenient to assess the accuracy of the model by comparing these predictions with measurements, and we do so in the Results section.

\subsection{Sensor modelling}
The optical irradiance at the sensor surface is converted to voltages and a digital output value using a sensor model. The Google Pixel 4a uses a Sony IMX363 sensor; we summarize the sensor specifications (e.g. pixel size, fill factor, sensor resolution) in \ref{appendix:sensor}. Some of these values were published by Sony and others were estimated in lab experiments described below. To perform these experiments, we used  OpenCamera\cite{opencamera}, a free and open-source software application that controls camera gain and exposure duration, to obtain the nearly unprocessed digital values from the sensor.  

\subsubsection{Sensor spectral quantum efficiency}
\begin{figure}[!ht]
  \includegraphics[width=1\columnwidth]{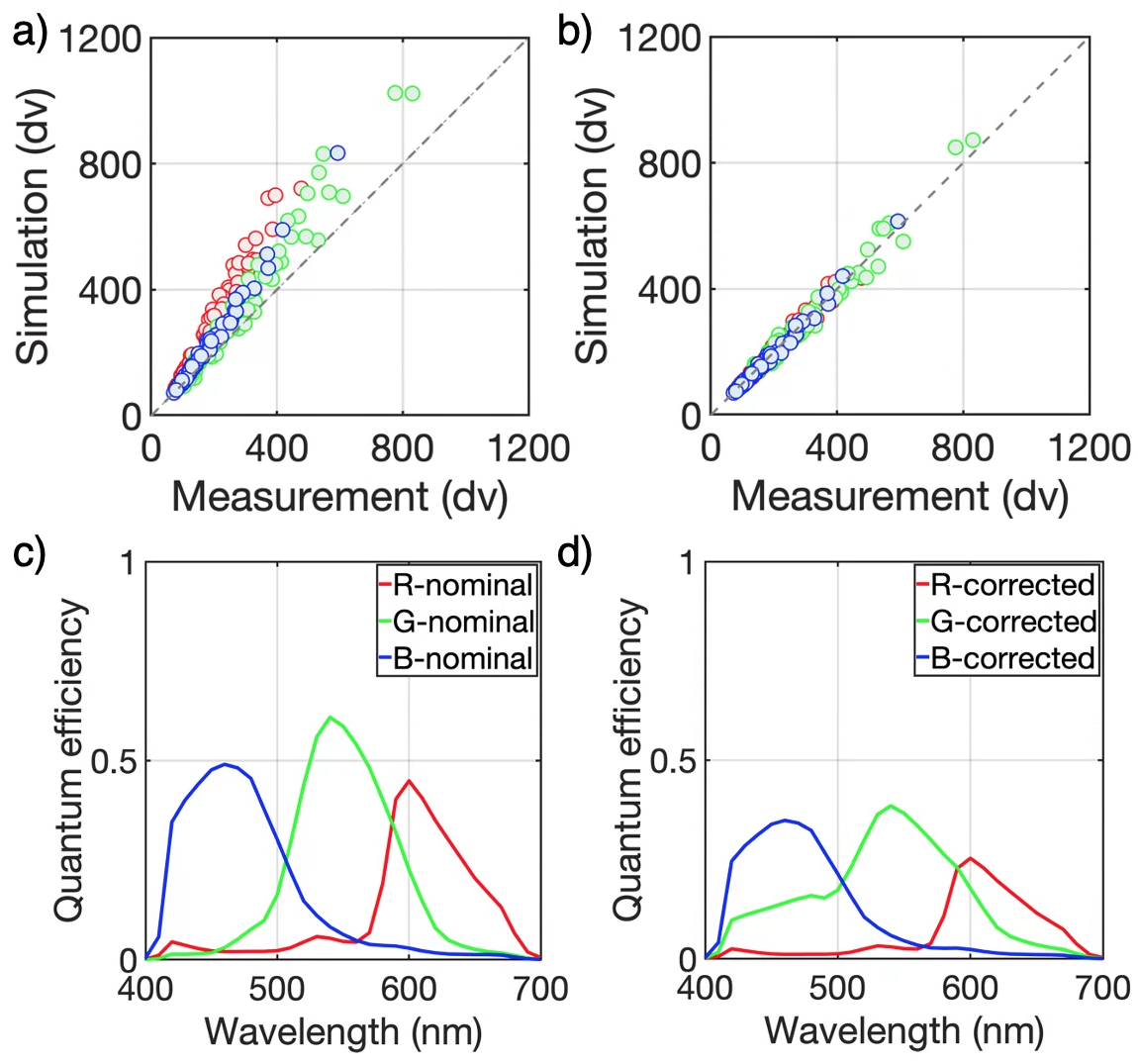}
  \caption{Sensor spectral quantum efficiency calibration. (a) Scatter plot comparing measured and simulated RGB values of the MCC based on nominal color filter quantum efficiency. (b) Scatter plot comparing measured and simulated RGB values after correction for cross-talk and channel gain. (c) Nominal color filter quantum efficiency and (d) quantum efficiency after correction.}
  \label{Fig-QEcalibration}
\end{figure}

 To calibrate the sensor spectral quantum efficiency (QE), we captured images of a Macbeth Color Checker (MCC) under three different illuminants. The spectral power distribution of the illuminants and the reflectance functions of the 24 MCC patches are plotted in Figure \ref{Fig-QEcalibration}(ab). As part of the estimation, it is necessary to reduce relative illumination (vignetting) effects. We did this by measuring and correcting for the relative illumination and by placing the MCC in the center of the camera image where the change in relative illumination is smaller. 
 
 We began the calibration by comparing the measured RGB values with those predicted by using the published sensor color channel quantum efficiencies (QE) and the transmissivity of a NIR filter as described in \cite{lyu2021validation}.  There is a substantial mismatch for this prediction which we attribute to optical and electrical cross-talk and channel gains. To account for these factors, we found a positive $3 \times 3$ matrix, $M$, that transforms the spectral QE.  The matrix was estimated by minimizing the least square error between the measurement ($r^\prime, g^\prime, b^\prime$) and the original prediction ($r, g, b$):

\begin{ceqn}
\begin{align}
\label{eq:transform}
\displaystyle{\min_{M}}~||\begin{bmatrix}r^\prime& g^\prime & b^\prime\end{bmatrix} - \begin{bmatrix} r& g& b\end{bmatrix} M||^2
\end{align}
\end{ceqn}
where $\begin{bmatrix} r^\prime, g^\prime, b^\prime \end{bmatrix}$ and $\begin{bmatrix} r,g,b \end{bmatrix}$ are  $72 \times 3$, (24 patches under three illuminants). We solve for $M$ subject to a non-negativity constraint
\begin{ceqn}
\begin{align}
\label{eq:condition}
M(i, j) \geq 0, 1 \leq i, j \leq 3
\end{align}
\end{ceqn}

The fitted matrix is:
\begin{ceqn}
\[M=
\begin{bmatrix}
0.5636 & 0.0807 & 0.0069\\
0 & 0.5917 & 0\\
0 & 0.2470 & 0.7098
\end{bmatrix}\]
\end{ceqn}

The diagonal entries of $M$ are channel gains. There is one significant off-diagonal value that represents cross-talk between the blue and green channels. After incorporating the correction matrix, the simulated sensor data match the measurement (Figure \ref{Fig-QEcalibration}d) within a few percent (8.9\%). The three spectral quantum efficiency curves for the  channels are shown before and after correction in the Figure \ref{Fig-QEcalibration} c-d, respectively.

%

\subsubsection{Sensor noise model}
We estimated the following noise sources \cite{emva-sensor-noise}: 
\begin{enumerate}
    \item \textbf{Photon noise (or shot noise)} is a natural consequence of the photoelectric effect. The number of electrons generated is Poisson distributed.
    \item \textbf{Dark current noise} is the leakage current within each pixel in the absence of light. The electrons are generated by thermal effects and crystallographic defects in the depletion region.
    \item \textbf{Dark signal non-uniformity} (DSNU) is random distribution of offset levels across pixels measured at short exposure durations in the absence of light.
    \item \textbf{Photo response non-uniformity} (PRNU) describes the variation of the gain (slope) when measuring the electrical signal as a function of incident photons.
    \item \textbf{Read noise} arises from electrical noise in the circuitry that reads out the pixel value.
    \item \textbf{Reset noise} arises from small fluctuations in the voltage level achieved when resetting the pixel prior to an acquisition.
\end{enumerate}
Some types of noise differ with each acquisition (temporal noise).  Other types of noise are variations across the sensor that remain consistent across acquisitions (fixed pattern).  Certain potential sources of variation, such as column fixed pattern noise, are very small and not considered in IMX363 sensor modelling.

\subsubsection{Combined noise model estimates}
The combined effect of different noise sources is present in the sensor digital values. The expected noise is signal-dependent, largely because the photon noise is signal-dependent. To evaluate the accuracy of the simulated noise we compared the standard deviation of the simulation and measurement in multiple regions that span a range of signal levels (Section \ref{val}).

\section{Results} \label{val}
\subsection{Qualitative appearance comparison}
\begin{figure}[!t]
  \includegraphics[width=1\columnwidth]{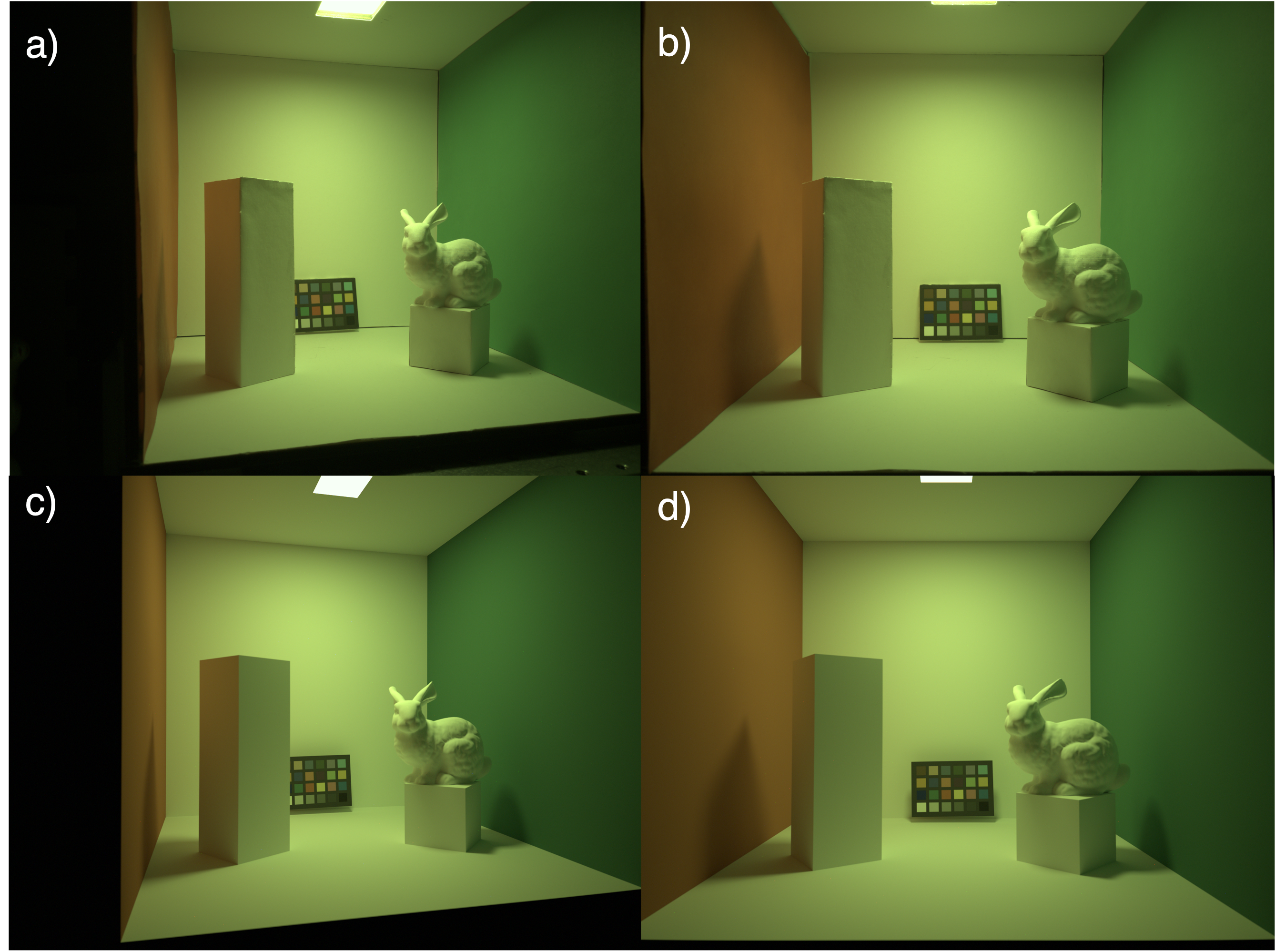}
  \caption{Qualitative comparisons between the simulation and measurement from two camera positions. (a) and (b) are measurements; (c) and (d) are simulations.}
  \label{Fig-qualitative}
\end{figure}

The simulated and measured sensor image data appear quite similar (Figure \ref{Fig-qualitative}). Both represent the high dynamic range of the scene. For example, the area light source is saturated while the shadows and corners are very dark. Light reflected from the colored walls can be seen on the sides of the cubes. 

We do not expect to observe a pixel-by-pixel match between the simulated and measured camera images. This is partly due to small differences in the positions of the objects and camera in the simulated and real scene. Furthermore, we modeled the scene illumination as a uniform area light. This approximation introduces some small differences. Hence, to characterize the simulation accuracy we compare a series of quantitative summary measures. In the next sections we evaluate relative illumination, optical blur, depth of field, chromatic channel responses and sensor noise.

\subsection{Relative illumination}
\begin{figure}[!t]
\begin{center}
  \includegraphics[width=0.7\columnwidth]{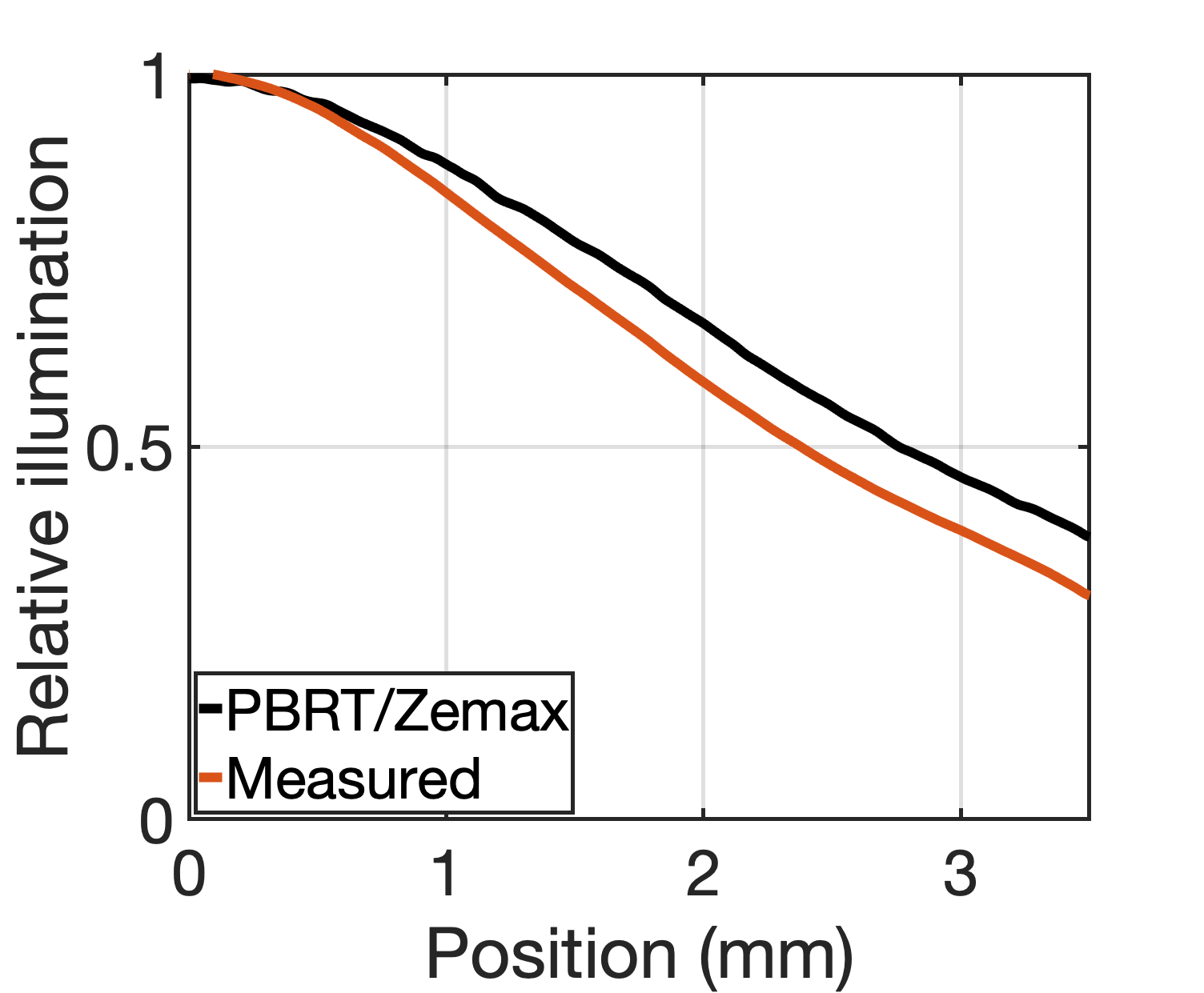}
  \caption{Comparison of simulated and measured relative illumination. The curves show the relative illumination from the center of the sensor to a 3.5 mm field height (the sensor diagonal is 7 mm).  The Zemax black box model relative illumination and ray-transfer prediction in PBRT are nearly identical (black). Neither closely matches the measured relative illumination (red), which falls-off more rapidly.}
  \label{Fig-lensvignettcomp}
\end{center}
\end{figure}

In many cell phone cameras, the optics introduce a substantial change in the relative illumination with field height. Using the Zemax black-box model of the Google Pixel 4a lens, we expect a fall-off from the center to the 3mm field height to be approximately 60\% (Figure \ref{Fig-lensvignettcomp}).  The relative illumination, calculated using the ray-transfer method described in \cite{Goossens-RTF}, matches the Zemax prediction very closely.  The measured relative illumination differs, falling off by approximately 70\%. The most likely cause of the differences is that we did not model the microlens array placed on the surface of nearly all commercial sensors. This information is not included in the Zemax black box model, and we did not have access to the microlens description.This effect is widely described in the literature \cite{SVS-Vistek_GmbH_undated-sd}. In subsequent calculations, we use the measured relative illumination as part of the simulation.

\subsection{MTF and depth of field}

%
%
\begin{figure}[!b]
\begin{center}
  \includegraphics[width=0.9\columnwidth]{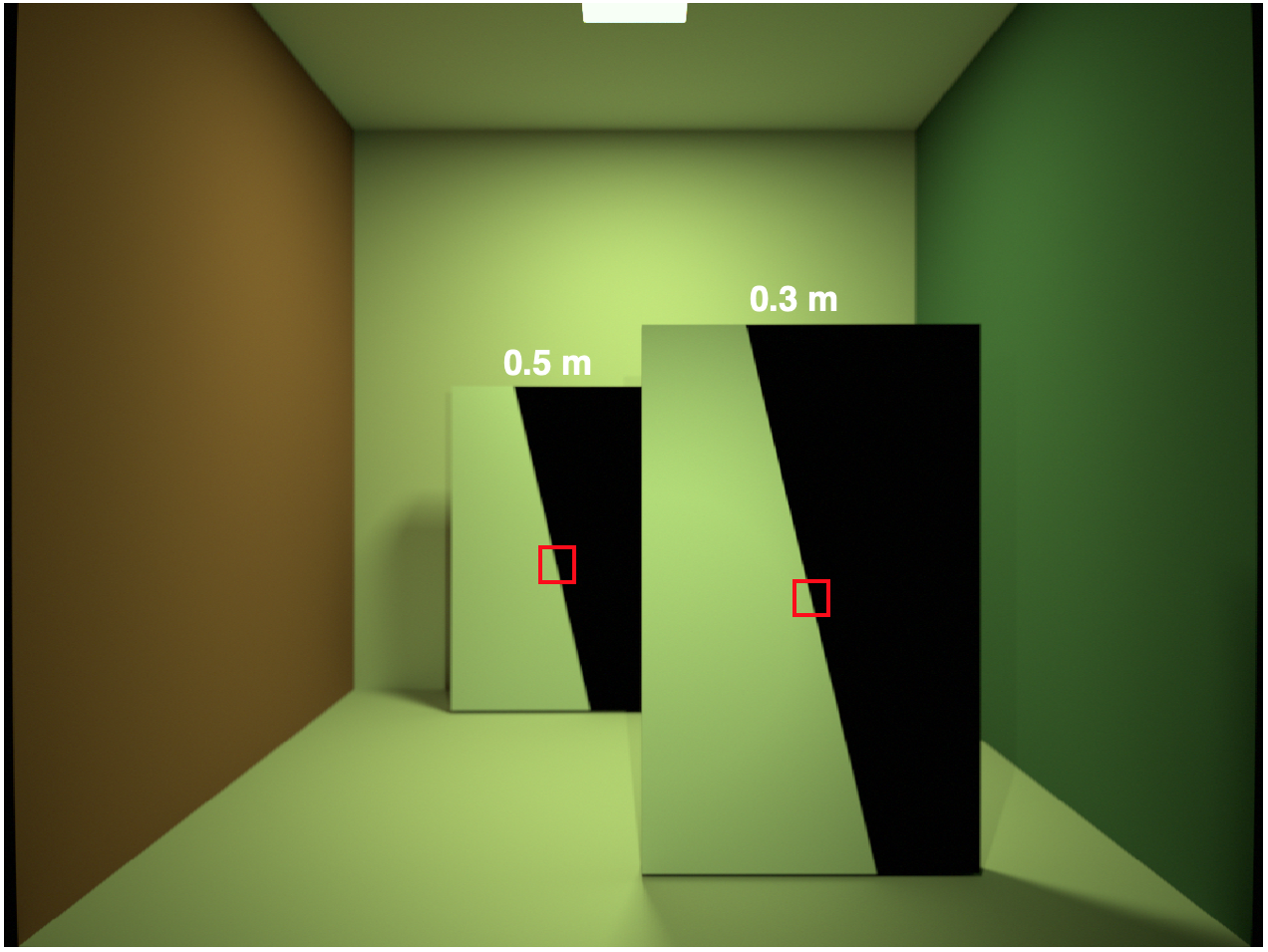}
  \caption{The line spread function (LSF) was estimated from a simulated image the contained two slanted edges. These were positioned at two different depths (0.3 m and 0.5 m from the camera position) and field heights (0.7 mm and 0.8 mm) that matched the distances and field heights of the slanted edge target that was captured with the Google Pixel 4a camera.  The LSF is estimated by selecting a small region of the slanted edge and using measurements from multiple rows to estimate a densely sampled step-edge response.  The derivative of the step-edge response is the LSF.  The modulation transfer function (MTF) is the magnitude of the Fourier Transform of the LSF.}
  \label{Fig-slantedbarpreview}
\end{center}
\end{figure}

The line spread function (LSF) measures how light from a thin line is spread across the sensor surface, serving as a summary measure of spatial blur.  The LSF is narrow for a line at the focal distance, broadening for lines placed closer or farther. We measured the LSF function of the camera using the methods defined by the ISO 12233 standard \cite{iso12233}. To validate the RTF, two slanted edge targets were placed in the Cornell box at 0.3 m and 0.5 m from camera position. Two images were acquired, one with the focal length set to each of these distances.  We expect that when one slanted edge target is in focus, the LSF for the other slanted edge target will be wider.  

\begin{figure}[!t]
\begin{center}
  \includegraphics[width=1\columnwidth]{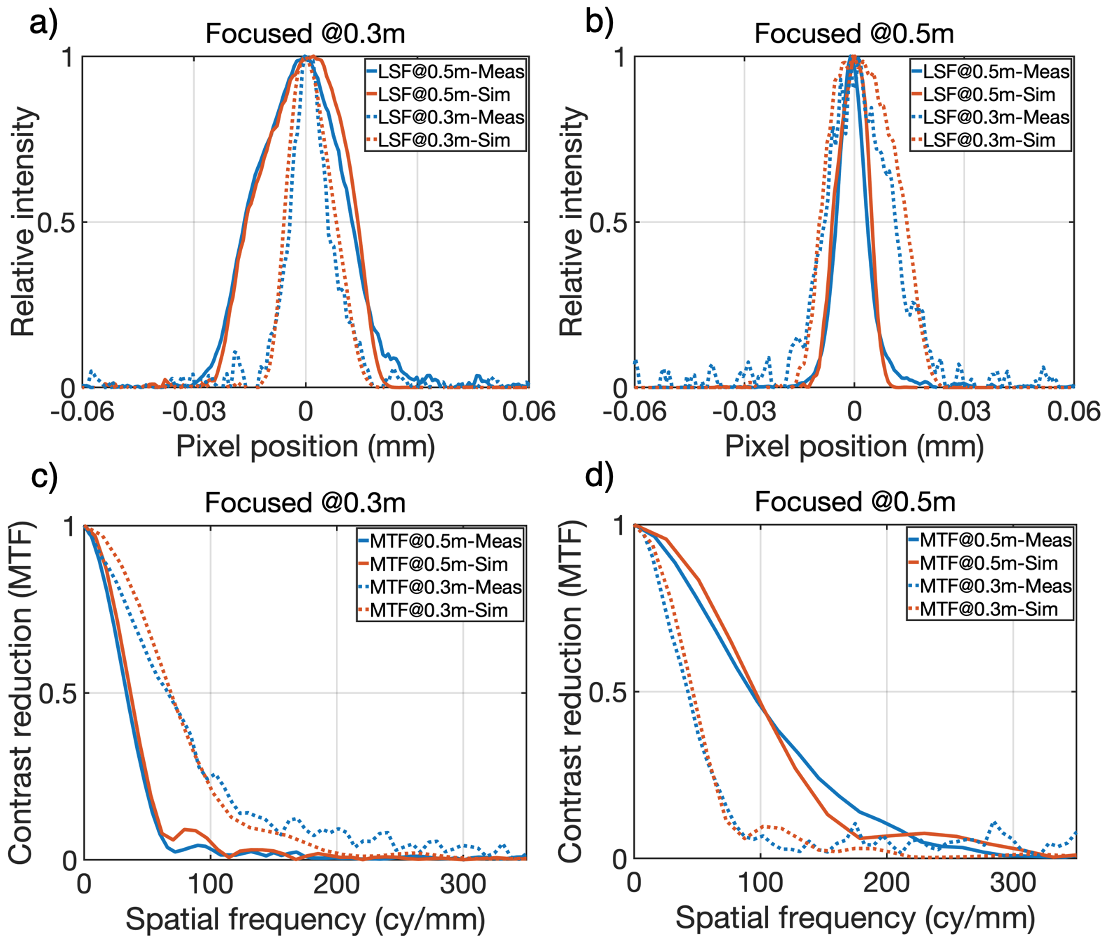}
  \caption{Comparison of the spatial blur and depth of field.  (a,b).  The measured (blue) and simulated (red) LSFs are compared. The curves show the LSFs for lines at two different distances.  Panel a shows data with a focal distance of 0.3m and Panel b with the focal distance of 0.5m. In each case the narrow LSF is for the line at the focal distance, and the broader LSF is for the line that is too near or too far. (c,d) The LSF data are transformed into the corresponding MTF curves where the broader MTF is for the line that is in focus. The measured and simulated lines were at slightly different field heights (0.7 mm and 0.8 mm), but this difference did not have a substantial impact on the measurements.}
  \label{Fig-lsfmtfdof}
\end{center}
\end{figure}

The simulated images in Figure \ref{Fig-slantedbarpreview} illustrate the geometry. The LSFs computed from the measured and simulated images are shown in Figure \ref{Fig-lsfmtfdof}.  Panel (a) compares the LSFs with the camera focused at 0.3 m and panel (b) compares the LSFs with the camera focused at 0.5 m. As expected, in both cases the LSF is broader when the line is away from the focal distance.  

To further compare the data, we transformed the LSFs to modulation transfer functions (MTFs) in panels (c) and (d). The MTF describes the reduction in image contrast as a function of image spatial frequency. The narrow LSF becomes a broader MTF.  Both the MTF and LSF measure the depth-dependence, which defines the depth of field.  This will vary with the lens and its aperture opening.  The Google Pixel 4a does not adjust the aperture, and for this reason we compared using focal distance.

\subsection{Sensor noise}
In this section we describe our evaluation of the sensor noise model. The model includes photon noise and several classes of temporal (reset, dark current, and read) and fixed pattern (DSNU, PRNU) noise. These sources contribute differently to the total noise, and some of these sources depend on the sensor irradiance level. 

\begin{figure}
\begin{center}
  \includegraphics[width=1\columnwidth]{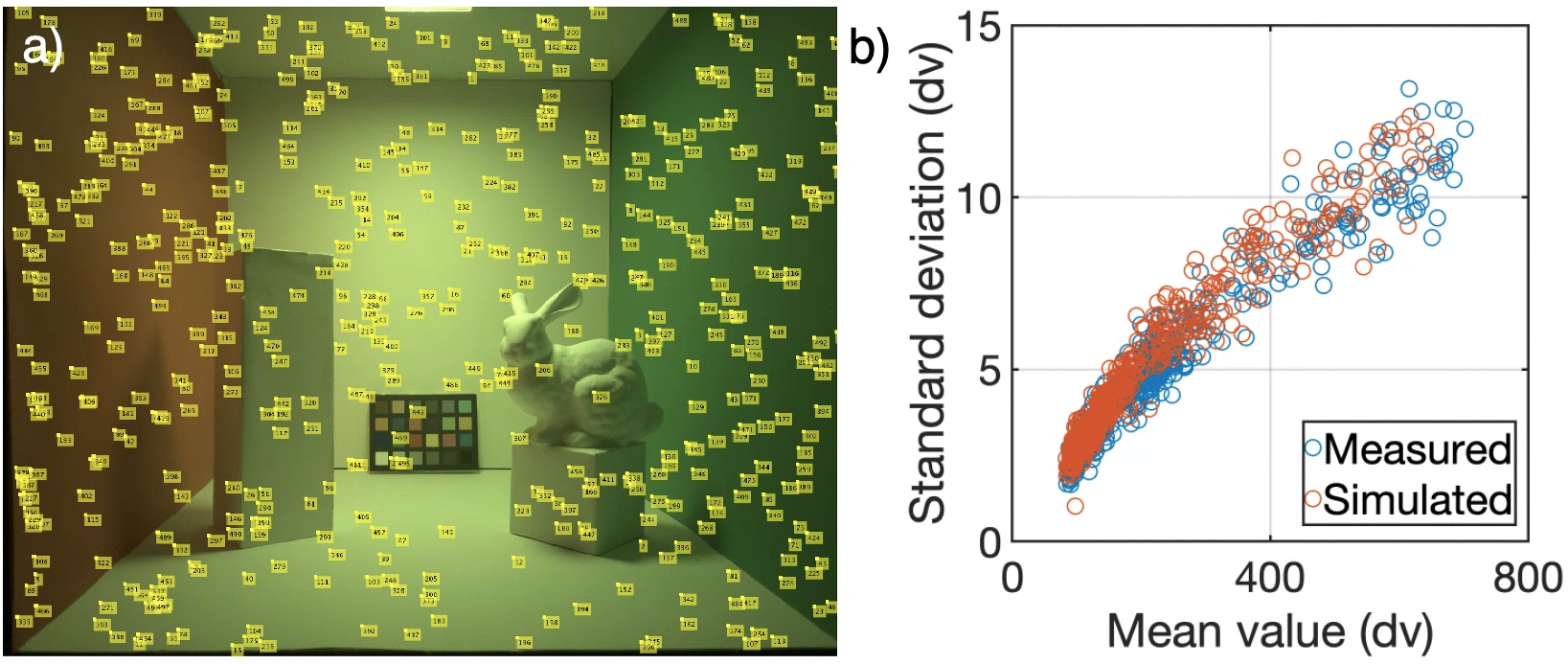}
  \caption{Sensor noise estimated in measurement and simulation. (a) We randomly tested many small ($10 \times 10$) image regions. We accepted a region for a noise estimate if it contained no dead pixels and the values were consistent with a uniform spectral irradiance.  The yellow boxes show the regions that were accepted.  See text for details.  (b) Within each uniform region, the standard deviation increases with mean signal level. The simulated data match well with the measurement.}
  \label{Fig-sensornoiseval}
\end{center}
\end{figure}

To assess the accuracy of the sensor noise model, we compared simulated and measured data from image regions with a wide range of digital levels. Specifically, we selected 500 uniform patches using criteria described in \ref{appendix:uniform}.  The regions are shown overlaid on a measured image in Figure \ref{Fig-sensornoiseval}a). The sensor noise, measured as the standard deviation of the green pixels, increases with the mean level (Figure \ref{Fig-sensornoiseval}b). The relationship between the standard deviation and the mean is similar whether estimated using the measured or simulated data. Hence, we conclude that the sensor noise model is a good approximation to the measured data.

\subsection{Inter-reflections}
\begin{figure*}[!ht]
\begin{center}
  \includegraphics[width=2\columnwidth]{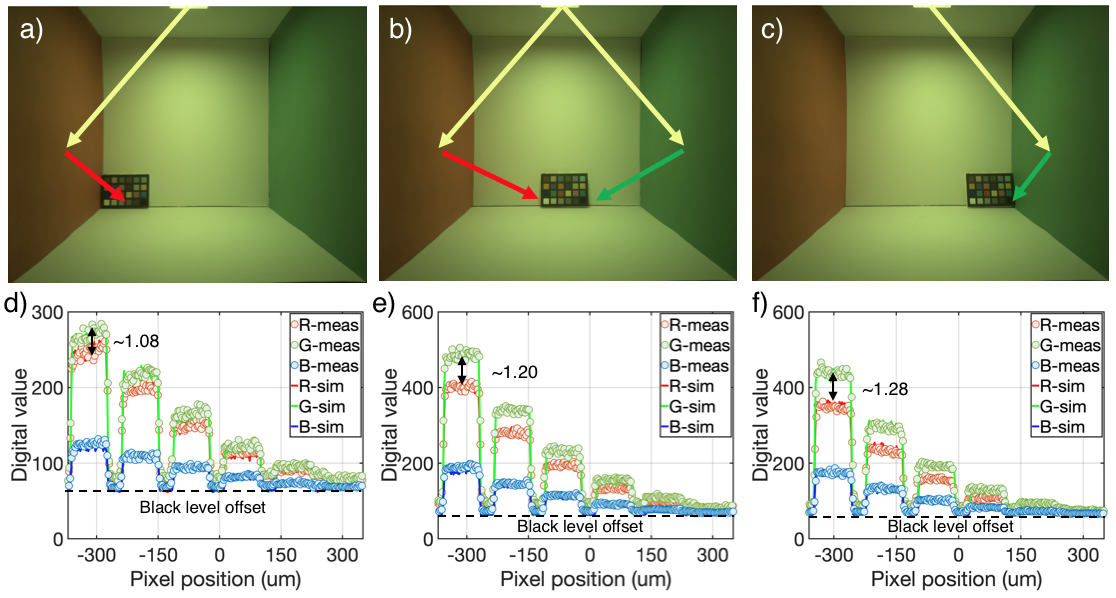}
  \caption{Comparisons of simulated and measured digital values in the complex scene including inter-reflections. (a-c) Acquired images of an MCC at three different positions along the rear wall.  The arrows show the likely paths for inter-reflections from the light source off the nearby walls onto the MCC.  The white rectangle highlights the position of the achromatic patches on the MCC.  (d-f)  The digital values from the Google Pixel 4a images (dashed lines) are compared with the simulated values (solid lines). See text for details.}
  \label{Fig-interreflection}
\end{center}
\end{figure*}
Next, we compare the simulated and measured digital values for an image that includes substantial inter-reflections.  We acquired and simulated images of the Cornell box that contained an MCC at the rear wall.  Figure \ref{Fig-interreflection})a-c show images that were measured when the MCC was positioned either closer to the left wall (red), right wall (green) or half-way in between. The graphs (Figure \ref{Fig-interreflection} d-f) below each image plot the digital values in a horizontal line passing through the achromatic series of the MCC in that image. The solid lines plot the digital values from the measured image and the points plot the digital values from the simulated image, with one free scalar adjustment. This adjustment was necessary because the angle of the simulated MCC is perfectly parallel to the rear wall, but the measured MCC was tilted slightly.  Hence, the amount of light scattered from the achromatic series was larger in the measured image than in the simulated image.  We therefore scaled all of the simulated values in the horizontal line passing through the achromatic series of the MCC by a single positive constant that brought the two sets of data into reasonable agreement.

There are several features worth noting when comparing the simulated and measured digital values.  First, notice that the ratio between the green:red channel ratio is close to one (1.08) at the left (red) wall and becomes significantly higher near the right (green) wall (1.28).  This change is due to the light reflected onto the achromatic patches from the nearby walls.  Second, the variance in the simulated and acquired data is quite similar at all levels.  Finally, there is a difference in the black line regions between the simulated and measured data.  This arises because the black lines in the simulated MCC were set to zero reflectance, but the black lines in the real MCC are slightly more reflective.  We left this imperfection in the simulation to show that it is possible to see such small mismatches. Note also that the lowest digital value contains the black level offset (64).

\subsection{Validation of chromatic channel and relative illumination}

\begin{figure}[!t]
\begin{center}
  \includegraphics[width=1\columnwidth]{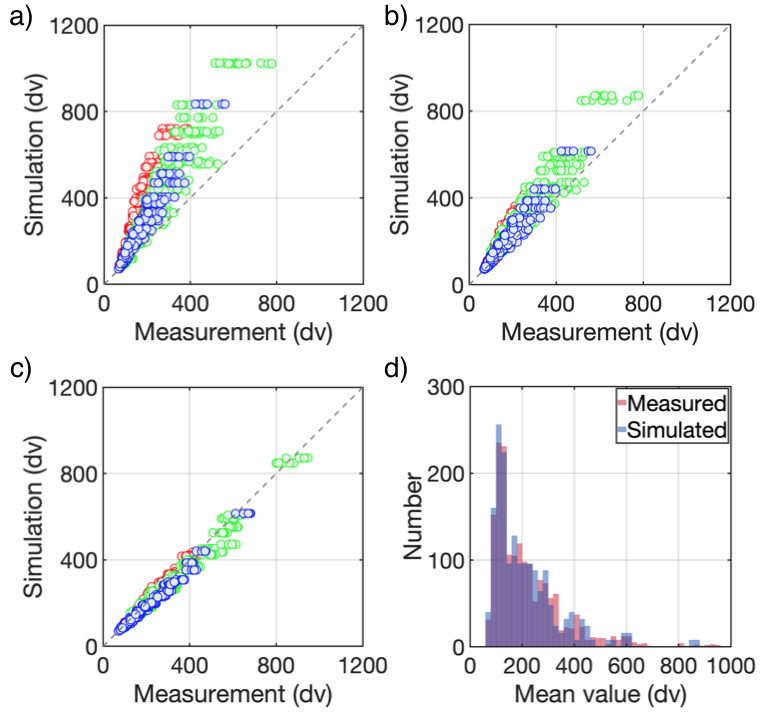}
  \caption{Validation of estimated color filter QE. The scatter plots compare the mean simulated and measured value for the 24 patches under three different illuminants. The color of each point is a measurement in the R, G or B channels. Panels a-c show these fits. (a) Without lens vignetting, channel gain, or cross-talk correction the fits are quite poor. (b) Correcting for lens vignetting alone improves the fit. (c) Correcting for lens vignetting, channel gain and cross-talk places all the points near the identity line. (d) The histogram shows the distribution of measured and simulated digital values; most of the comparisons are based on relatively small digital values.}
  \label{Fig-colorfilterheight}
\end{center}
\end{figure}

Finally, we performed one more assessment of the accuracy of the sensor QE by comparing simulated and measured R, G and B digital values of the Macbeth color checker. This validation was performed with a miniature version of the MCC in the center of a Gretag light booth.  The booth has 3 different lights nominally labeled "illuminant A", "CWF" and "day".  For each light, we captured images using the Google Pixel 4a camera in 8 different positions, for a total of 576 color patch comparisons (3 illuminants x 8 positions x 24 color patches).  Moving the camera while keeping the position of the MCC fixed places the MCC at different field heights and preserves the same illumination on the MCC. To correctly simulate these sensor responses, we must have an accurate model of both the color channels and the relative illumination. 

We simulated RGB values in several steps. First, we multiplied the spectral reflectance of each of the 24 MCC color patches with the spectral power of the illumination and the spectral quantum efficiencies (QE) of the Google Pixel 4a camera. Second, we corrected for relative illumination by dividing the measured RGB values by the measured relative illumination (see Figure \ref{Fig-lensvignettcomp}). Third, we used the conversion gain (see Appendix \ref{appendix:conversion}) to calculate the sensor RGB digital values. Finally, we added the digital black level (64) to the simulated values.

Figure \ref{Fig-colorfilterheight} compares 1728 measured and simulated mean RGB values (576 patches with 3 values) at several steps in the calculation.  We first compare values without accounting for relative illumination and using the nominal sensor QE functions (Figure \ref{Fig-QEcalibration}a). Next, we compare the measured and simulated RGB values after correcting for sensor gain and cross-talk (see Figure \ref{Fig-QEcalibration}b). Finally, we make the same comparison also accounting for the effect of relative illumination (Figure \ref{Fig-colorfilterheight}c). Each additional step improves the fit between simulated and measured data. The histogram panel \ref{Fig-colorfilterheight}d illustrates the distribution of digital values in the data set.  


\section{Discussion} \label{Discussion}
We assessed simulation accuracy using a series of quantitative measurements: Our end-to-end image systems simulation matches the camera data in many ways. Simulations of the optical blur and depth of field agreed with the measurements to within a few percent (Figures \ref{Fig-lsfmtfdof} and \ref{Fig-colorfilterheight}). Simulations of the spatial pattern of sensor responses within a complex scene  (Figure \ref{Fig-interreflection}) and the sensor noise (Figure \ref{Fig-sensornoiseval}) are also in close agreement with the measurements. The quantitative agreement between the simulated and measured camera image data are a strong validation of the end-to-end simulation methodology.

The validation is based on a three-dimensional scene that includes significant shadows and inter-reflections.  The light scattered from the colored walls are a substantial factor in illuminating the sides of the cubes. The cubes cast significant shadows on the walls.  All of these features are congruent in the simulated and measured camera images.

Modeling commercial camera optics is a long-standing challenge in image systems simulation.  Goossens et al. \cite{Goossens-RTF} developed the ray-transfer method that enables us to model a Google Pixel 4a lens even though the specific design is unknown. We found that key properties determined by the optics - relative illumination, optical blur and depth of field - can be matched even though the specific lens components are not specified. These results show that the ray transfer function obtained from a Zemax black model for the Google Pixel 4a camera can be substituted for a lens model in PBRT.

Prior work had already established the accuracy of linear plus noise sensor models for planar scenes with simple lighting \cite{Farrell2008-sc,Chen2009-bj}. Image system simulation validation based on 3D scenes with fluorescent materials and sensors were described in \cite{Lyu2021-ae}. The tests here extend those validations by comparing the R, G and B digital values in simulated and measured camera images of the MCC under multiple light sources and different camera positions. In both of these cases, using different cameras, we found it necessary to estimate the channel gains and crosstalk. 

After including these calibrations, the mean pixel values in the simulated and measured camera images of the MCC lie near the identity line (Figure \ref{Fig-colorfilterheight}c). The agreement between the mean pixel values in the simulated and measured camera images of the MCC test chart, illuminated with three different lights, confirms the general accuracy of the sensor model. The standard deviation of the points around the identity line, however, is larger than expected based on sensor and photon noise alone (Figure \ref{Fig-sensornoiseval}b). To explain this additional variance may require more precise accounting for factors such as the relative illumination. 



Accurate sensor modeling is necessary for designing novel sensor architectures. Soft-prototyping using realistic scene data can shorten development time by enabling designers to select imaging components and avoid mistakes. Image systems simulation also has value in developing subsequent signal processing and machine learning algorithms.  Consider, for example, that machine learning developers for driving applications are using synthetic camera images to augment data sets and create specific training scenarios (e.g., \cite{omniversevalidation-Kamel2021,anyverse-mainpage,waymo-simulationcity}. Each of those vendors emphasize different types of validations for their simulations. A contribution of this work is the focus on properties of the optics and sensor that are beyond what we have found in the literature.

Correctly modeling the optics and sensor properties, including pixel size and color filter arrays, is important for many applications. For example, neural networks trained on physically-based simulations of camera images can detect cars in real camera image data almost as well as neural networks trained on the real camera image data \cite{liu2020neural}. Furthermore, neural networks trained on camera images generated by physically-based simulations performed better than neural networks trained on camera images generated by raster-based graphics or by ray-traced graphics rendering methods that did not include the correct optics and sensor modeling \cite{Liu2019-ds, liu2020neural,Liu2021-mr,liu2021isetauto}. 

The results of our study support the use of physically based end-to-end image systems simulations to create large data sets that are automatically and accurately labeled for training neural networks. The results also support using the simulations to assess how hardware changes will impact system performance. These analyses can be helpful in fields apart from consumer photography, such as medical diagnosis, driving, and robotics.

\section{Future work} \label{connfuture}
The results that we report in this paper give us confidence in our ability to accurately simulate a camera imaging complex, natural scenes. Some of the limitations in this work are a good target for future work.

For example, ISETCam and ISET3d include the ability to simulate microlens arrays.  We did not have this information, and the missing information produced a difference between the simulated and measured relative illumination. We are investigating methods for accounting for the microlens array when this information is not provided.

Related, we are simulating sensors with multiple photosites beneath each microlens to measure the light field.  A particularly simple form of these sensors, the dual pixel architecture, is already in wide use for setting focus. A useful extension of the analysis here is to simulate dual pixel and light field sensors.  These simulations will help us understand how much information can reliably be extracted from these devices.

We are also exploring other types of imaging systems, such as time-of-flight and gated-CMOS cameras.  Our simulation of such devices has been enabled by the ability to use ray tracing to calculate the path length of each ray from the sensor into the scene, including the effect of optical elements and filters that are in the light path.  Further work that accounts for the participating media on the light path (fog, smoke), and the bidirectional scattering distribution functions of the materials at the relevant wavelengths, will be helpful in the design of such systems and in synthesizing realistic data sets.  Such data sets can be used to extract information about the scene using different algorithms, or simply to create labeled training data for machine learning applications.

As we noted, an important goal for this project is to build trust in these simulations. We accomplish this by assessing the accuracy of our simulations and by making our image systems simulation software open-source and freely available.  We continue to update our software repositories with new data and functions that are part of our ongoing research so that others can check our work and build on it. By making these soft-prototyping tools easily available, we hope to advance the design of imaging sensors for future imaging applications.

\appendices

\section{Render configuration} \label{appendix:render configurations}
The optical images of the Cornell Box scene were rendered using PBRT-V3 \cite{Pharr2016-yb} with the RTF modifications. The images were rendered at the same spatial sampling resolution as the Sony sensor (3024 $\times$ 4032).  The rendering parameters were set to 6 bounces and 3072 rays per pixel.  These parameters were chosen to reduce the rendering noise to a very small level. Rendered on a CPU with 40 cores, the rendering time was 15 hours. Experiments with PBRT-V4 and a 3080 Nvidia GPU suggest that the rendering time for the same scene will be approximately 1.5 hours.

\section{Sensor parameters} \label{appendix:sensor}
Sensor parameters used for simulation are in Appendix Table \ref{tab:sensor}.  The pixel size, fill factor, well capacity, voltage swing, conversion gain, black level offset and quantization are provided by the manufacturer. We also estimated conversion gain (see Appendix \ref{appendix:conversion}).

We estimated the dark signal nonuniformity (DSNU), photoresponse nonuniformity (PRNU), dark voltage and read noise from lab measurements. Specifically we used an integrating sphere to produce images of uniform intensity.  We then acquired a set of images for a range of integration times.  We fit the measurements from a collection of different exposure times to estimate these quantities (see methods in \cite{Farrell2012-ma}).

\begin{table}[!h]
\caption{Sony IMX363 sensor specification.}
\label{tab:sensor}
\begin{center}       
\begin{tabular}{|p{0.25\columnwidth}|p{0.30\columnwidth}|p{0.30\columnwidth}|}
\hline
\textbf{Properties} & \textbf{Parameters} & \textbf{Values (units)} \\ \hline\hline
\multirow{2}{*} {Geometric} & Pixel Size & [1.4, 1.4] ($\mu$m) \\
                            & Fill Factor & 100 ($\%$) \\ \hline
\multirow{6}{*} {Electronics} & Well Capacity & 6000 ($e^-$) \\
                         & Voltage Swing & 0.4591 (volts) \\
                         & Conversion Gain & $ 0.1707 (dv/e^-$) \\
                         & Analog Gain & 1 \\
                         & Black Level Offset & 64 (dv) \\
                         & Quantization Method & 10 (bit) \\ \hline      
\multirow{4}{*} {\makecell[l]{Noise Sources \\ @Analog gain=1}} & DSNU & 0.038 (mV) \\
                          & PRNU & 0.54 (\%) \\
                          & Dark Voltage & 0.02 (mV/sec) \\
                          & Read Noise & 0.226 (mV) \\ \hline                         

\end{tabular}
\end{center}
\end{table}

\section{Estimating conversion gain} \label{appendix:conversion}
We estimate the conversion gain by analyzing the pixel values in an image of a bright scene, where we expect shot noise to be dominant. In the ideal case, the number of excited electrons, which is the signal $\Tilde{S}$, follows a Poisson distribution \cite{Janesick2007-sc}:
\begin{equation}
    \Tilde{S} \sim \text{Poisson}(\mu) .
\end{equation}
The mean and variance are equal. In practice, the observed signal also depends on the photo-response nonuniformity (PRNU) and dark signal nouniformity (DSNU).  The measured DSNU is very small, and so we do not include it in the calculations.  The PRNU $\Tilde{P}$ follows a normal distribution:
\begin{equation}
    \Tilde{P} \sim 1 + \mathcal{N}(0, \sigma_{\text{prnu}}).
\end{equation}
The observed signal $\Tilde{O}$ will be the product of these two independent random variables: 
\begin{equation}
    \Tilde{O} = \Tilde{S} \Tilde{P}
\end{equation}
The expected value of the number of observed electrons can be expressed in terms of the mean and variance of these two random variables
\begin{equation}
\label{eq:obsmean}
\begin{split}
\text{E}(\Tilde{O}) = \text{E}[\Tilde{S}\Tilde{P}]
          = \text{E}(\Tilde{S})\text{E}(\Tilde{P})
          = \mu
\end{split}
\end{equation}

The variance of the observed signal $\text{V}(\Tilde{O})$ is:
\begin{flalign}
\label{eq:obsvariance}
\begin{split}
\text{V}(\Tilde{O}) = &\text{V}[\Tilde{S}\Tilde{P})]\\
            = & [\text{E}(\Tilde{S})]^2 \text{V}(\Tilde{P}) + [\text{E}(\Tilde{P})]^2 \text{V}(\Tilde{S}) + \text{V}(\Tilde{S})\text{V}(\Tilde{P}) \\
            = &\mu^2\sigma_{\text{prnu}}^2+\mu+\mu\sigma_{\text{prnu}}^2
\end{split}
\end{flalign}

Since we can only measure digital values DV the conversion gain $\alpha$ is defined as converting electrons directly to digital value that has unit of $(dv/e^-)$:
\begin{equation}
\begin{split}
\text{DV}= \alpha \Tilde{O}
\end{split}
\end{equation}

The relationships of expectation and variance between digital value and observed signal are:

\begin{flalign}
\begin{split}
\text{E}(\text{DV})= & \alpha \text{E}(\Tilde{O}) \\
\text{V}(\text{DV})= & \alpha^2 \text{V}(\Tilde{O})
\end{split}
\end{flalign}

Substituting in Equations \ref{eq:obsmean} and \ref{eq:obsvariance} we have

\begin{flalign}
\begin{split}
\text{V}(DV) = & \alpha^2 \text{V}(\Tilde{O}) \\
      = & \alpha^2 (\mu^2\sigma_{\text{prnu}}^2 + \mu + \mu\sigma_{\text{prnu}}^2)\\
      = & \alpha^2 (\frac{[\text{E}(DV)]^2 \sigma_{\text{prnu}}^2}{\alpha^2} + \frac{\text{E}(DV)}{\alpha} + \frac{\text{E}(DV)}{\alpha}\sigma_{\text{prnu}}^2) \\
      = & [\text{E}(DV)]^2 \sigma_{\text{prnu}}^2 + \alpha \text{E}(DV)(1 + \sigma_{\text{prnu}}^2)
\end{split}
\end{flalign}

Finally, rearranging terms we have a solution for $\alpha$ in terms of the measured digital values from a uniform scene and the measured PRNU.

\begin{equation}
\begin{split}
 \alpha = 
 \frac{\textup{V}(\text{DV})-[\text{E}(\text{DV})]^2\sigma_{\text{prnu}}^2}{\text{E}(\text{DV})(1+\sigma_{\text{prnu}}^2)} .
 \end{split}
\end{equation}

\begin{figure}[!t]
\begin{center}
  \includegraphics[width=1\columnwidth]{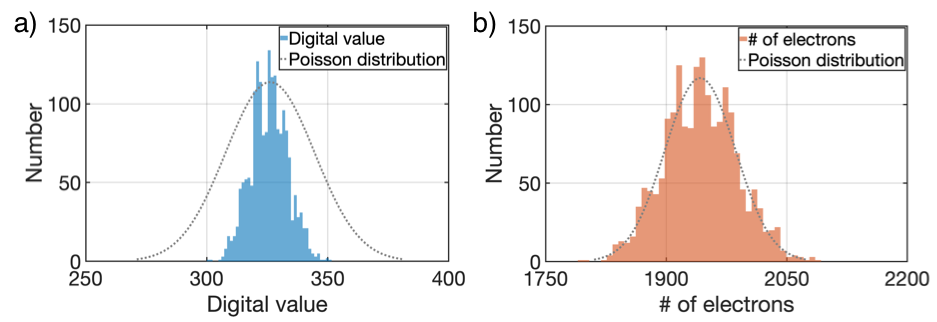}
  \caption{Histogram of (a) digital values.  The mean is equal to 320.7 and the variance is equal to 57.8. (b) After scaling by the inverse of the conversion gain we can obtain an estimate of the number of electrons.  In this case, the distribution mean and variance are more nearly equal as expected in Equation (6). This is consistent with the expected Poisson character of the light.}
  \label{Fig-conversiongain}
\end{center}
\end{figure}

To estimate the mean and variance of the digital values of a bright uniform scene, we used 1500 pixel values taken from 15 images acquired using an integrating sphere (see Appendix \ref{appendix:sensor}). We took  10 $\times$ 10 crops near the center of each image; this minimized the influence of lens vignetting. The estimated conversion gain 0.1677 $dv/e^-$ is close to the value provided by the manufacturer (0.1707 $dv/e^-$), differing by only 1.8\%. Figure \ref{Fig-conversiongain} shows the histogram of the digital values (panel a) and estimated number of electrons (b). The distribution of digital values is very different from Poisson. After applying the estimated conversion gain to  estimated the number of electrons, the distribution is close to Poisson, as expected.

\section{Uniform region identification} \label{appendix:uniform}
To estimate sensor noise, we chose regions in the image with uniform response levels. We randomly sampled a large number of $10 \times 10$ square regions in both the simulated and measured images. For each region we extracted the 50 green channel responses and selected regions based on an analysis of the digital values.

We rejected any region from measured images that contained an obviously 'dead' pixel or significant image non-uniformity. To accomplish this, we first converted the median value of the green pixel responses to electrons, using the conversion gain (See Appendix \ref{appendix:conversion}). In the absence of sensor noise, a uniform region would have a variance equal to the mean, as predicted by the Poisson distribution of photon noise. We rejected any region that had a pixel value more than 3 standard deviations from the mean.  We performed the same analysis on the simulated data, though in that case we attribute the rejection to rendering noise.

Second, we defined a criteria for non-uniformity. We held out 20\% of green values and fit the remaining values with a 2\textsuperscript{nd} order spatial polynomial. We then compared the root mean squared error (RMSE) of the held out data to the polynomial fit and to a constant value set to the mean of the green pixel values. We rejected a region as non-uniform if the RMSE of the fit to the constant was larger (2\% greater) than the RMSE of the polynomial fit. 

\section*{Acknowledgement} \label{acknowledge}
We thank Krithin Kripakaran for constructing the Cornell Box.  We thank Max Furth and Eric Tang for building a model of the Cornell Box in Cinema 4D.  We thank Zhenyi Liu, Henryk Blasinski and David Cardinal for their contributions in developing and supporting the ISET3d code and also for many helpful discussions. We thank Gordon Wan, Jamyuen Ko, Guangxun Liao, Bonnie Tseng and Ricardo Motta for their advice, support and encouragement for this project.

\bibliographystyle{ieeetr}
\bibliography{CornellBoxIEEE}

\begin{IEEEbiography}[{\includegraphics[width=1in,height=1.25in,clip,keepaspectratio]{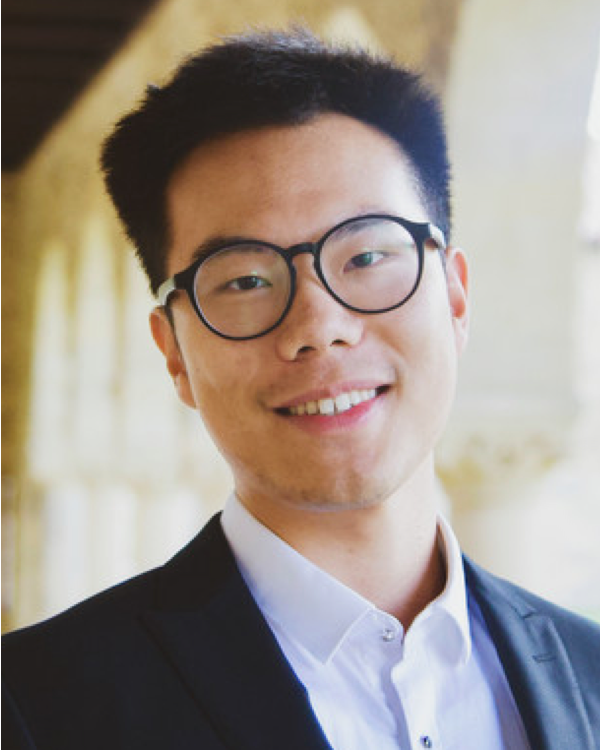}}]{Zheng Lyu} is a now pursing Ph.D. in the Department of Electrical Engineering at Stanford University, advised by Prof. Brian Wandell and Dr. Joyce Farrell. His research interests focus on physically based imaging system full pipeline simulation for consumer photography and medical imaging.
\end{IEEEbiography}
\vskip -2\baselineskip plus -1fil
\begin{IEEEbiography}[{\includegraphics[width=1in,height=1.25in,clip,keepaspectratio]{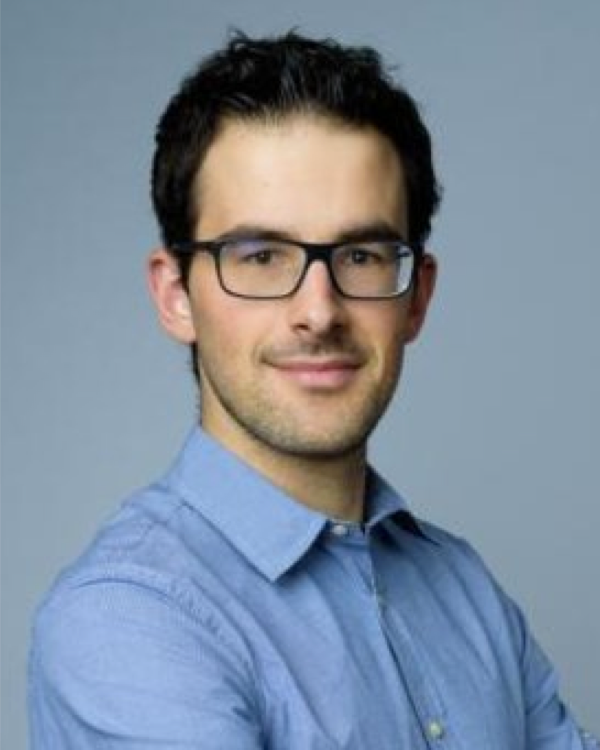}}]{Thomas Goossens} received his Ph.D. at KU Leuven, working in coordination with scientists at IMEC.  He is currently a postdoctoral fellow at Stanford University working on the physics of patterned photonic structures and camera simulation. He is the holder of a B.A.E.F fellowship. 
\end{IEEEbiography}
\vskip -2\baselineskip plus -1fil
\begin{IEEEbiography}[{\includegraphics[width=1in,height=1.25in,clip,keepaspectratio]{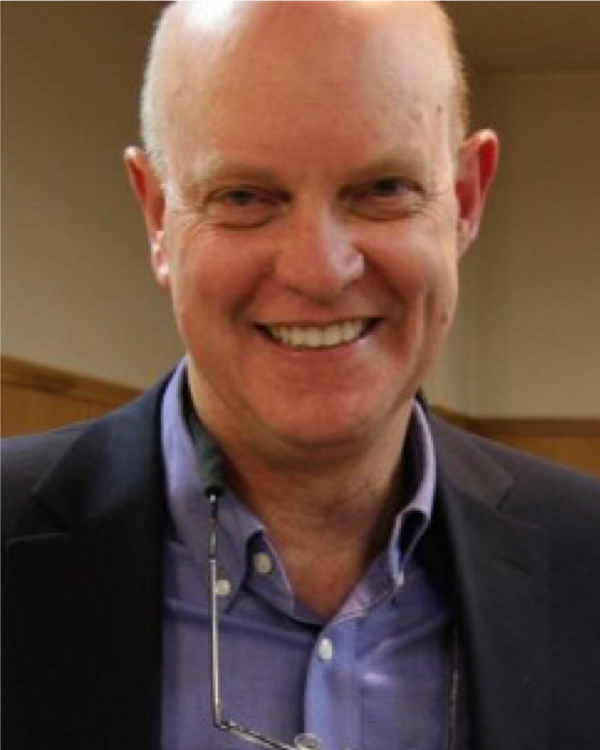}}]{BRIAN A. WANDELL}is the first Isaac and Madeline Stein Family Professor. He joined the Stanford
Psychology Faculty, in 1979, and is a member,
by courtesy, of Electrical Engineering, Ophthalmology, and the Graduate School of Education.
He is the Director of the Stanford’s Center for
Cognitive and Neurobiological Imaging and the
Deputy Director of the Stanford’s Neurosciences
Institute. His research interests include vision science, spanning topics from visual disorders, reading development in children, to digital imaging devices, and algorithms for
both magnetic resonance imaging and digital imaging.
\end{IEEEbiography}
\vskip -2\baselineskip plus -1fil
\begin{IEEEbiography}[{\includegraphics[width=1in,height=1.25in,clip,keepaspectratio]{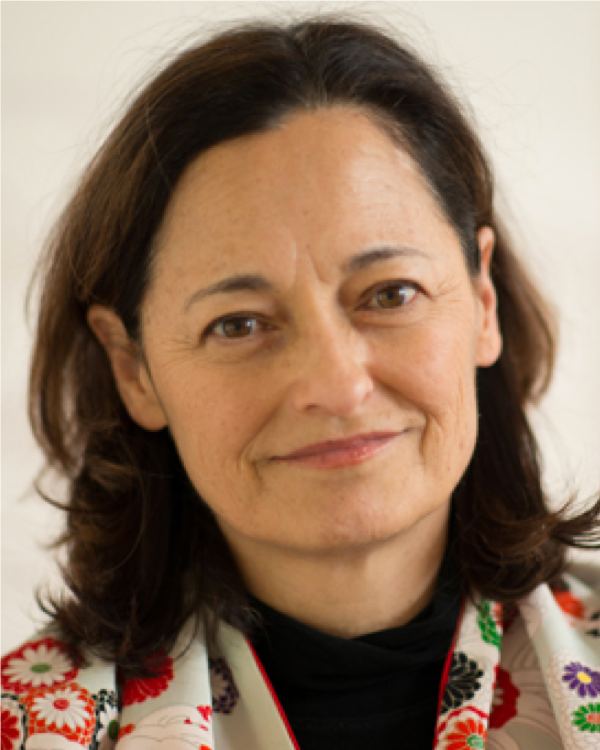}}]{JOYCE FARRELL}is a Senior Research Engineer
and a Lecturer with the Department of Electrical Engineering, Stanford University. She is also
the Executive Director of the Stanford Center
for Image Systems Engineering. Dr. Farrell has more than 20 years of research and professional experience  working at a variety of companies and institutions, including the NASA Ames Research Center, New York University, the Xerox Palo Alto Research Center, Hewlett Packard Laboratories and Shutterfly
\end{IEEEbiography}

\EOD
\end{document}